\documentclass[11pt]{article}%
\usepackage{amsfonts}
\usepackage{amsmath}
\usepackage{amssymb}
\usepackage{graphicx}%
\providecommand{\U}[1]{\protect\rule{.1in}{.1in}}
\newtheorem{theorem}{Theorem}

\newenvironment{proof}[1][Proof]{\noindent\textbf{#1.} }{\ \rule{0.5em}{0.5em}}
\begin{document}

\title{Observing Quantum Trajectories: From Mott's Problem to Quantum Zeno Effect and Back}
\author{Maurice de Gosson\thanks{University of Vienna, Faculty of Mathematics (NuHAG)
Oskar-Morgenstern-Platz 1, 1090 Vienna.}
\and Basil Hiley\thanks{Physics Department, University College, London, Gower
Street, London WC1E 6BT and TPRU, Birkbeck, University of London, Malet
Street, London WC1E 7HX.}
\and Eliahu Cohen\thanks{H.H. Wills Physics Laboratory, University of Bristol,
Tyndall Avenue, Bristol, BS8 1TL, U.K.}}
\maketitle

\begin{abstract}
The experimental results of Kocsis \emph{et al.}, Mahler \emph{et al.} and the
proposed experiments of Morley \emph{et al.} show that it is possible to
construct ``trajectories'' in interference regions in a two-slit
interferometer. These results call for a
theoretical re-appraisal of the notion of a ``quantum trajectory'' first
introduced by Dirac and in
the present paper we re-examine this notion from the Bohm perspective based on
Hamiltonian flows. In particular, we examine the short-time propagator and the
role that the quantum potential plays in determining the form of these
trajectories. These trajectories differ from those produced in a typical
particle tracker and the key to this difference lies in the \emph{active}
suppression of the quantum potential necessary to produce Mott-type
trajectories. We show, using a rigorous mathematical argument, how the active
suppression of this potential arises. Finally we discuss in detail how this
suppression also accounts for the quantum Zeno effect.

\end{abstract}

\section{Introduction}


There has been a revival of interest in the question of whether any meaning
can be given to the notion of a particle trajectory in the quantum domain
where field theory has already been so successful. However, the question
originally raised by Kemmer~\cite{nk39} remains, namely, how do we
discuss the limiting process through which the particle is identified by the
track it leaves in a tracking device like a bubble chamber. To construct a
trajectory, we need the notion of a local momentum, a notion that has been
assumed to be ruled out by the uncertainty principle. However, the uncertainty
principle is about a \emph{simultaneous measurement} of the position and
momentum and cannot answer the question as to whether or not the quantum
particle actually \emph{has} a simultaneous value of its position and
momentum. A further question remains, namely, that if a local momentum exists,
how do we measure it?

Wiseman~\cite{hw03} was one of the first to point out that the weak value of
the
momentum operator is the local momentum or the Bohm momentum in the Bohm
approach. (See also Hiley~\cite{bh12} for a wider perspective.)
Duck, Stevenson and Sudarshan~\cite{idpsgs89} have shown how weak values can
be measured in what are called weak measurements, a result
that immediately opens up the possibility of an experimental investigating the
trajectories such as those calculated by Philippidis, Dewdney and
Hiley~\cite{cpcdbh79} in the interference region of a two-slit
interferometer.
Indeed, the local momentum (essentially Poynting's vector~\cite{kbab13}) of
the electromagnetic field has already been measured in the experiments of
Koscis \emph{et al.}~\cite{skbb11} and Mahler \emph{et al.}~\cite{dmas15} who
used a quantum dot to generate a weak intensity field which was then passed
through a two-slit system. Measurements of the local momentum were then used
to construct what they called \textquotedblleft average photon trajectories''.
In this sense they are returning to the notion of a ``quantum trajectory''
first introduced by Dirac~\cite{pd45}.

Although these flow lines have some resemblance to the trajectories of
Philippidis \emph{et al.}~\cite{cpcdbh79} referred to above, they cannot be
compared directly because the latter are calculated using the Schr\"{o}dinger
equation, whereas photons are excitations of the electro-magnetic field. A
treatment of the field approach from the Bohmian point of view has been given
by Bohm, Hiley and Kaloyerou~\cite{dbbhpk87} and by Kaloyerou~\cite{pk94}.
Thus it is not clear that photons can be considered to be travelling along
trajectories. However Morley, Edmonds and Barker~\cite{jmjepb16} are now
carrying out a similar experiment using, instead, argon atoms with an aim to
construct trajectories which can be directly compared with the theoretically
predicted ones.

In light of this background, we explore the relation between the trajectories
determined in a general interference region and the trajectories that are seen
in a particle tracker more closely using the Hamiltonian flow method developed
by de Gosson~\cite{ICP} which, in turn, clarifies the Bohm approach as
discussed in Bohm and Hiley~\cite{BoHi}.
This latter approach centres on the real part of the Schr\"{o}dinger equation
under polar decomposition of the wave function, which takes the form
\begin{eqnarray}
\frac{\partial S^{\psi}}{\partial t}+\frac{(\nabla S^{\psi})^{2}}%
{2m}+V(x)+Q^{\Psi}(x,t)=0	\label{eq:QHJ}
\end{eqnarray}

where $Q^{\Psi}(x,t)$ is the quantum potential defined by
\[
Q^{\Psi}(x,t)=-\frac{\hbar^{2}}{2m}\frac{\nabla^{2}R^{\psi}(x,t)}{R^{\psi
}(x,t)}.
\]
This equation, the quantum Hamilton-Jacobi equation, like its classical
counterpart, enables us to calculate an ensemble of trajectories once we are
given a solution of the Schr\"{o}dinger equation for the experimental
situation under investigation. Clearly, the key to the different forms of
trajectories lies in the appearance of the quantum potential. We will show how
this potential arises naturally in the method of Hamiltonian flows and study
its behaviour in more detail.

As two of us \cite{gohi2} have already shown for the particular case of a
point source, the quantum motion of the particle always reduces to the
classical Hamiltonian trajectory for short times. Thus, in this case the
non-appearance of the quantum potential will guarantee classical behaviour. In
this paper we generalize our theory to the case of arbitrary initial
conditions. In this way we provide a rigorous proof of how the Bohmian
approach explains this phenomenon without appealing to any wavefunction collapse.

Our investigations show that the key to the appearance of the classical
behaviour is the suppression of the quantum potential. Indeed we find that all
quantum phenomena arise from the presence of this term in the real part of the
Schr\"{o}dinger equation. Therefore, its suppression will inhibit quantum
transitions and this is what is required to explain the quantum Zeno effect.

A few words regarding the structure of this paper.

\begin{itemize}
\item In Section \ref{sec1} we review the notion of short-time propagators for
the Schr\"{o}dinger equation using previous works of ours
\cite{ICP,gohi2,gohi3} and of Makri and Miller \cite{makmil1,makmil2}; the
main result is that one can replace the Van Vleck propagator -- which is
accurate to order $\Delta t^{2}$ -- with what we call the Kerner--Sutcliffe
propagator \cite{Kerner}, which is much easier to use in explicit calculations
since it does not require the evaluation of the exact action integral, but
only of simply computable $\Delta t^{2}$ approximation thereof;

\item In Section \ref{sec2} we begin by revisiting the Bohmian theory of
quantum motion, which reinstates the notion of trajectory in quantum
mechanics; we emphasize the Hamiltonian nature of this motion, which has
already been highlighted by Holland \cite{Holland1,Holland2}. We thereafter
utilize the results from Section \ref{sec1} to show that to a very good
approximation (i.e. to order $\Delta t^{2}$) these Bohmian trajectories are
\emph{classical} Hamiltonian trajectories.

\item In Section \ref{sec3} we employ our results to the study of the quantum
Zeno effect for an arbitrary system of particles, using an extension to
time-dependent flows of the usual Lie--Trotter formula. We apply our results
to the Mott problem.
\end{itemize}

We have highlighted the main mathematical results by dignifying them as
\textquotedblleft Theorems\textquotedblright\ rather than hiding them in a
morass of calculations.

We will be working in position and momentum coordinates $x=(x_{1},...,x_{n})$
and $p=(p_{1},...,p_{n})$ referring to a N-body system; the corresponding
phase space is thus $2n$ dimensional. We recall Landau's \textquotedblleft big
O\textquotedblright\ notation: given two functions $f(t)$ and $g(t)$ the
relation $f(t)-g(t)=\mathcal{O}(t^{k})$ (for $t\rightarrow0)$ means that there
exists a constant $C$ such that $|f(t)-g(t)|\leq C|t^{k}|$.

\section{Short-Time Action and Approximate Propagators\label{sec1}}

We begin by listing the assumptions we make on the quantum Hamiltonian. We
thereafter shortly review the well-known approximation to the quantum
mechanical propagator essentially due to Van Vleck. We then show that one can,
with benefit, replace the latter with a simpler approximation, whose
construction is based on exact asymptotic expressions of the action integral.

\subsection{Basic assumptions\label{sec11}}

Let $\Psi$ be a solution of the multi-particle Schr\"{o}dinger equation%
\begin{equation}
i\hbar\frac{\partial\Psi}{\partial t}(x,t)=\left[  \sum_{j=1}^{n}\frac
{-\hbar^{2}}{2m_{j}}\frac{\partial^{2}}{\partial x_{j}^{2}}+V(x)\right]
\Psi(x,t). \label{schrodeq}%
\end{equation}
We will assume that given a square integrable initial value $\Psi
(x,t_{0})=\Psi_{0}(x)$ this solution exists and is unique. We moreover assume
that if $\Psi_{0}(x)$ is infinitely differentiable in the position variables
$x_{1},...,x_{n}$ so is $\Psi(x,t)$ for every time $t$ in some (usually small)
interval $I_{T}=[-T,T]$. This assumption allows the quantum potential $Q(x,t)$
and its gradient (the quantum force, up to the sign) to be defined for $t\in
I_{T}$.

More precisely, we assume that the conditions for Stone's theorem that
establishes a one-to-one correspondence between self-adjoint operators on the
Hilbert space $L^{2}(\mathbb{R}^{n})$ and one-parameter families of $(U_{t})$
unitary operators are satisfied. This requires that the quantum Hamiltonian
\begin{equation}
\widehat{H}=\sum_{j=1}^{n}\frac{-\hbar^{2}}{2m_{j}}\frac{\partial^{2}%
}{\partial x_{j}^{2}}+V(x) \label{hamop}%
\end{equation}
be \textit{essentially self-adjoint} on $L^{2}(\mathbb{R}^{n})$, in which case
the family $(U_{t})$ is formally given by $U_{t}=e^{-i\widehat{H}t/\hslash}$
(see for instance Reed and Simon \cite{reedsimon}). It is well-known
(following Kato \cite{Kato}) that this is the case for the majority of
physically interesting potentials $V(x)$, including Coulomb-type potentials of
the form $r^{-m}$ for $m<3/2$. The essential point in Kato's proof is that we
should have
\begin{equation}
V\Psi\in L^{2}(\mathbb{R}^{n})\text{ for all }\Psi(x)=P(x)e^{-|x|^{2}/2}
\label{Kato}%
\end{equation}
where $P(x)=P(x_{1},...,x_{n})$ is a polynomial in position coordinates.
Notice that Kato's condition also trivially holds when the Hamiltonian
function is a quadratic polynomial in the position variables (one can produce
in this case explicit formulas for $U_{t}$ using the metaplectic
representation \cite{ICP,Birk}). More generally (\ref{Kato}) will hold for all
potential functions that are continuously differentiable to all orders in the
$x_{1},...,x_{n}$ variables, and for which there exists, for every multi-index
$\alpha\in\mathbb{N}^{n}$ real constants $C_{\alpha}>0$ and $m_{\alpha}$ such
that%
\begin{equation}
|\partial_{x}^{\alpha}V(x)|\leq C_{\alpha}(1+|x|)^{m_{\alpha}}.
\label{conditionv}%
\end{equation}

Condition (\ref{conditionv}) can be relaxed in various ways; for instance one
can assume that the potential is only continuously differentiable up to a
finite order ensuring the existence of the quantum potential and its derivatives.

Defining for two arbitrary times $t,t^{\prime}\in I_{T}$ the operator
$U_{t,t^{\prime}}=U_{t}U_{-t^{\prime}}$ on $L^{2}(\mathbb{R}^{n})$ we
have:\medskip

(1) $U_{t,t^{\prime}}U_{t^{\prime},t^{\prime\prime}}=U_{t,t^{\prime\prime}}$
and $U_{t,t}=I_{\mathrm{d}}$ for all $t,t^{\prime},t^{\prime\prime}\in I_{T}$;

(2) $U_{t,t^{\prime}}$ is strongly continuous in $L^{2}(\mathbb{R}^{n})$ with
respect to $t,t^{\prime}$;

(3) $\Psi(x,t)=U_{t,t_{0}}\Psi_{0}(x)$ is a solution of (\ref{schrodeq}) for
every $\Psi_{0}\in L^{2}(\mathbb{R}^{n})$;

(4) $U_{t,t^{\prime}}$ maps $L^{2}(\mathbb{R}^{n})\cap C^{\infty}%
(\mathbb{R}^{n})$ into itself for all $t,t^{\prime}\in I_{T}$.\medskip

Notice that it follows by induction from the groupoid property $U_{t,t^{\prime
}}U_{t^{\prime},t^{\prime\prime}}=U_{t,t^{\prime\prime}}$ that for every
integer $N\geq1$
\begin{equation}
U_{t,t_{0}}=U_{t,t_{N-1}}U_{t_{N-1},t_{N-2}}\cdot\cdot\cdot U_{t_{1},t_{0}}%
\end{equation}
hence we also have
\begin{equation}
U_{t,t_{0}}=\lim_{N\rightarrow\infty}(U_{t,t_{N-1}}U_{t_{N-1},t_{N-2}}%
\cdot\cdot\cdot U_{t_{1},t_{0}})
\end{equation}
and this identity also holds if we replace in the right-hand side the
operators $U_{t_{j+1},t_{j}}$ with second-order approximations $\overline
{U}_{t_{j+1},t_{j}}=U_{t_{j+1},t_{j}}+\mathcal{O}((t_{j+1}-t_{j})^{2})$:%
\begin{equation}
U_{t,t_{0}}=\lim_{N\rightarrow\infty}(\overline{U}_{t,t_{N-1}}U_{t_{N-1}%
,t_{N-2}}\cdot\cdot\cdot\overline{U}_{t_{1},t_{0}}).
\end{equation}
This follows from the Lie--Trotter--Kato theory (Chorin \emph{et al.}
\cite{chorinetal}), and will be used in Section \ref{secbon} (formula
(\ref{Trotter}).

It follows from Schwartz's kernel theorem that the conditions (1)--(3) ensure
the existence of a distribution $K=K(x,x^{\prime},t,t^{\prime})$ (the
\textquotedblleft propagator\textquotedblright)\ such that
\[
\Psi(x,t)=\int K(x,x^{\prime},t,t^{\prime})\Psi(x^{\prime},t^{\prime}%
)d^{n}x^{\prime};
\]
the property $U_{t,t}=I_{\mathrm{d}}$ is equivalent\ to $K(x,x^{\prime
},t,t)=\delta(x-x^{\prime})$; it follows that for fixed $x^{\prime}$ and
$t^{\prime}$ the function $K(x,x^{\prime},t,t^{\prime})$ is a solution of
Schr\"{o}dinger equation (\ref{schrodeq}) with initial condition $\Psi
^{\prime}(x,t^{\prime})=\delta(x-x^{\prime})$. Physically, it represents the
wavefunction of a point source located at $x^{\prime}$ at the initial time
$t^{\prime}$.

Here is a (very) short guide to the existing literature complementing the
discussion above: Doi \cite{doi} studies the regularity problem of the
solutions to (\ref{schrodeq}) from the point of view of energy estimates.
Yajima \cite{ya87} states some precise condition, weaker than
(\ref{conditionv}), guaranteeing the existence of the propagator for
potentials generalising the Coulomb case. Leforestier \emph{et al.} \cite{num}
discuss in depth various approximate propagation schemes for very general
potentials which can easily be numerically implemented.

\subsection{The Van Vleck Propagator}

We will denote the initial time by $t_{0}$ and write $\Delta t=t-t_{0}$.

Consider the Schr\"{o}dinger equation (\ref{schrodeq}) with initial value
$\Psi(x,t_{0})=\Psi_{0}(x)$; we assume that $\Psi_{0}(x)$ is infinitely
differentiable and belongs to $L^{2}(\mathbb{R}^{n})$. We will denote by
\[
K(x,x_{0},t,t_{0})=\langle x|U_{t,t_{0}}|x_{0}\rangle=\langle x|e^{-\frac
{i}{\hbar}\widehat{H}(t-t_{0})}|x_{0}\rangle
\]
the corresponding exact propagator. The solution of (\ref{schrodeq}) at time
$t$ is thus given by the formula
\begin{equation}
\Psi(x,t)=\int K(x,x_{0},t,t_{0})\Psi_{0}(x_{0})d^{n}x_{0}.\label{kernel}%
\end{equation}
The condition $\lim_{t\rightarrow t_{0}}\Psi(x,t)=\Psi_{0}(x)$ implies that we
must have
\begin{equation}
\lim_{t\rightarrow0}K(x,x_{0},t,t_{0})=\delta(x-x_{0});\label{cond1}%
\end{equation}
since $K$ satisfies Schr\"{o}dinger's equation for fixed $x_{0}$ and $t_{0}$.
This can be viewed as the wavefunction corresponding to a point source located
at $x_{0}$ at time $t=t_{0}$.

It is well-known (see Gutzwiller's discussion in \cite{Gutz}, \S 1.6 and 12.5;
also \cite{ICP,MF,schulman}) that for short times an approximate propagator is
given by Van Vleck's formula%
\begin{equation}
\widetilde{K}(x,x_{0},t,t_{0})=\left(  \tfrac{1}{2\pi i\hbar}\right)
^{n/2}\sqrt{\rho(x,x_{0},t,t_{0})}e^{\frac{i}{\hbar}S(x,x_{0},t,t_{0})}
\label{vanvleck}%
\end{equation}
where the function
\begin{equation}
S(x,x_{0},t,t_{0})=\int_{t_{0}}^{t}\left(
{\textstyle\sum_{j=1}^{n}}
\tfrac{1}{2}m_{j}\dot{x}_{j}(s)^{2}-V(x(s))\right)  ds \label{actionex}%
\end{equation}
is the action along the classical trajectory leading from $x_{0}$ at time
$t=t_{0}$ to $x$ at time $t$ (there is no sum over different classical
trajectories because only one trajectory contributes in the limit
$t\rightarrow t_{0}$), and%
\begin{equation}
\rho(x,x_{0},t,t_{0})=\det\left(  -\frac{\partial^{2}S(x,x_{0},t,t_{0}%
)}{\partial x_{j}\partial x_{0,k}}\right)  _{1\leq j,k\leq n} \label{rho}%
\end{equation}
is the Van Vleck density of trajectories; the argument of the square root in
(\ref{vanvleck}) is chosen so that the initial condition (\ref{cond1}) is
satisfied. A \textit{caveat}: the Van Vleck propagator is frequently used in
semiclassical mechanics \cite{Gutz}, it is however not a \textquotedblleft
semiclassical\textquotedblright\ object \textit{per se}: it is genuinely an
approximation to the exact propagator for small values of $\Delta t=t-t_{0}$
-- not just in the limit $\hbar\rightarrow0$. In fact (see for instance
\cite{ICP}, Lemma 241), setting%
\[
\widetilde{\Psi}(x,t)=\int\widetilde{K}(x,x_{0},t,t_{0})\Psi_{0}(x_{0}%
)d^{n}x_{0}%
\]
we have%
\begin{equation}
\Psi(x,t_{0}+\Delta t)-\widetilde{\Psi}(x,t_{0}+\Delta t)=\mathcal{O}(\Delta
t^{2}) \label{estpsitilde}%
\end{equation}
and hence, in particular,
\begin{equation}
K(x,x_{0},t_{0}+\Delta t,t_{0})-\widetilde{K}(x,x_{0},t_{0}+\Delta
t,t_{0})=\mathcal{O}(\Delta t^{2}) \label{kk}%
\end{equation}
for small values of $\Delta t=t-t_{0}$.

\subsection{The Kerner and Sutcliffe propagator\label{secbon}}

While the van Vleck propagator (\ref{vanvleck}) is accurate to order $\Delta
t^{2}$, it is, however, difficult to calculate in practice since it involves
the action integral (\ref{actionex}), whose determination in principle
requires the resolution of Hamilton's equations for the classical Hamiltonian
$H$, which is, outside a few trivial cases, a quite cumbersome task. We
therefore construct another simpler approximation to same order $\Delta t^{2}$
of the exact propagator. For this purpose we begin by remarking that the exact
action integral (\ref{actionex}) is approximated by the function
\begin{equation}
\overline{S}(x,x_{0},t,t_{0})=\sum_{j=1}^{n}m_{j}\frac{(x_{j}-x_{0,j})^{2}%
}{2(t-t_{0})}-\overline{V}(x,x_{0})(t-t_{0}) \label{sbar}%
\end{equation}
where $\overline{V}(x,x_{0})$ is the average of the potential $V$ along the
line segment $[x_{0},x]:$
\begin{equation}
\overline{V}(x,x_{0})=\int_{0}^{1}V(\tau x+(1-\tau)x_{0})d\tau. \label{vbar}%
\end{equation}
In fact, one proves \cite{ICP,makmil1,makmil2} that for short times $\Delta t
$ we have
\begin{equation}
S(x,x_{0},t_{0}+\Delta t,t_{0})-\overline{S}(x,x_{0},t_{0}+\Delta
t,t_{0})=\mathcal{O}(\Delta t^{2}). \label{estimateICP}%
\end{equation}
For this, it suffices to remark that the function $S=S(x,x_{0},t,t_{0})$
satisfies the Hamilton--Jacobi equation%
\begin{equation}
\frac{\partial S}{\partial t}+\sum_{j=1}^{n}\frac{1}{2m_{j}}\left(
\frac{\partial S}{\partial x_{j}}\right)  ^{2}+V(x)=0 \label{hamjac}%
\end{equation}
and one thereafter looks for an asymptotic solution%
\begin{equation}
S(x,x_{0},t,t_{0})=\frac{1}{t-t_{0}}S_{0}(x,x_{0})+S_{1}(x,x_{0}%
)(t-t_{0})+\mathcal{O}((t-t_{0})^{2}). \label{add}%
\end{equation}
Insertion in (\ref{hamjac}) then leads to
\[
S_{0}(x,x_{0})=\sum_{j=1}^{n}m_{j}\frac{(x_{j}-x_{0,j})^{2}}{2}%
\]
(which is the free generating function) and $S_{1}(x,x_{0})=-\overline
{V}(x,x_{0})$ hence (\ref{sbar}). We emphasize that this approximation to the
action is very different from the \textquotedblleft
midpoint-rules\textquotedblright\ commonly used in the theory of the Feynman
path integral \cite{schulman}, which only yield, at best, $\mathcal{O}(\Delta
t)$ approximations (see the discussions in
\cite{ICP,gohi2,gohi3,makmil1,makmil2}). We moreover point out that this
procedure can be used to obtain approximations of $S(x,x_{0},t,t_{0})$ to
arbitrary order $\mathcal{O}(\Delta t^{N})$ by adding terms $S_{j}%
(x,x_{0})(t-t_{0})^{j}$ ($j=2,...,N$) to the asymptotic sum (\ref{add}) and
determining the functions $S_{j}(x,x_{0})$ by successive quadratures (see
Makri and Miller \cite{makmil1,makmil2} for explicit calculations).

The next step consists of setting
\begin{equation}
\overline{H}(x,x_{0})=\sum_{j=1}^{n}\frac{p_{j}^{2}}{2m_{j}}+\overline
{V}(x,x_{0}) \label{hbar}%
\end{equation}
where $\overline{V}(x,x_{0})$ is the averaged potential (\ref{vbar}). One then
sets%
\begin{equation}
\overline{K}(x,x_{0},t,t_{0})=\left(  \tfrac{1}{2\pi\hbar}\right)  ^{n}\int
e^{\tfrac{i}{\hbar}(p(x-x_{0})-\overline{H}(x,x_{0},p)(t-t_{0}))}d^{n}p;
\label{KSpropagator}%
\end{equation}
using the theory of Fresnel integrals, it is easy to show after a few
calculations that%
\begin{equation}
\overline{K}(x,x_{0},t,t_{0})=\left(  \frac{m_{1}\cdot\cdot\cdot m_{n}}{2\pi
i\hbar(t-t_{0})}\right)  ^{n/2}e^{\frac{i}{\hbar}\overline{S}(x,x_{0}%
,t,t_{0})} \label{kbarfree}%
\end{equation}
where $\overline{S}(x,x_{0},t)$ is given by (\ref{sbar}), and the argument of
the square root is chosen so that $\lim_{t\rightarrow0}\overline{K}%
(x,x_{0},t)=\delta(x-x_{0})$. Using the short-time estimate (\ref{estimateICP}%
) one then shows that%
\begin{equation}
K(x,x_{0},t_{0}+\Delta t,t_{0})-\overline{K}(x,x_{0},t_{0}+\Delta
t,t_{0})=\mathcal{O}(\Delta t^{2}) \label{essentiel}%
\end{equation}
for small $\Delta t$. Defining the approximate wavefunction%
\begin{equation}
\overline{\Psi}(x,t)=\int\overline{K}(x,x_{0},t,t_{0})\Psi_{0}(x_{0}%
)d^{n}x_{0} \label{psibardef}%
\end{equation}
it follows that
\begin{equation}
\overline{\Psi}(x,t_{0}+\Delta t)-\Psi(x,t_{0}+\Delta t)=\mathcal{O}(\Delta
t^{2}) \label{psipsibar}%
\end{equation}
for small values of $\Delta t$. The approximations above are, in fact, of
order $O(\Delta t^{2})$; in view of formula (\ref{kk}), they are thus as good,
for all practical purposes, as the Van Vleck approximation $\widetilde
{K}(x,x_{0},t)$ considered in the previous section; the propagator
$\overline{K}(x,x_{0},t,t_{0})$ can, however, be easily calculated since it
only involves the determination of the integral (\ref{vbar}) giving the
average value of the potential along the straight line leading from $x_{0}$ to
$x$.

We notice that the propagator $\overline{K}(x,x_{0},t,t_{0})$ was postulated
by Garrod \cite{Garrod}, as well as Kerner and Sutcliffe \cite{Kerner}, though
they failed to prove the estimates (\ref{psibardef}) and (\ref{psipsibar}).
For details we refer to de Gosson's recent works \cite{go16-1,go16-2}.

We mention that the fact that $\overline{K}(x,x_{0},t,t_{0})$ is an
$\mathcal{O}(\Delta t^{2})$ approximation to the exact propagator implies that
we can construct the exact evolution operator for Schr\"{o}dinger's equation
by a \textquotedblleft time-slicing\textquotedblright\ process \cite{ICP}. In
fact, defining the unitary operator $U_{t,t_{0}}$ by
\[
U_{t,t_{0}}\Psi_{0}(x)=\int K(x,x_{0},t,t_{0})\Psi_{0}(x_{0})d^{n}x_{0}%
\]
(it is the exact evolution operator for the Schr\"{o}dinger equation) and its
(non-unitary) approximation
\[
\overline{U}_{t,t_{0}}\Psi_{0}(x)=\int\overline{K}(x,x_{0},t,t_{0})\Psi
_{0}(x_{0})d^{n}x_{0}%
\]
we have, in view of the estimate (\ref{psipsibar}) and the discussion in
Section \ref{sec11}),
\begin{equation}
U_{t,t_{0}}=\lim_{N\rightarrow\infty}(\overline{U}_{t,t_{N-1}}\overline
{U}_{t_{N-1},t_{N-2}}\cdot\cdot\cdot\overline{U}_{t_{1},t_{0}})
\label{Trotter}%
\end{equation}
where $t_{0}<t_{1}<t_{2}<\cdot\cdot\cdot<t_{N-1}<t$ is a subdivision of the
interval $[t_{0},t]$ such that $|t_{j+1}-t_{j}|<(t-t_{0})/N$. This can be
proven using telescopic sums and the Lie--Trotter--Kato formula (see the
Appendix B of \cite{ICP}, or, for more details, Chorin \textit{et al.}.
\cite{chorinetal} and Nelson \cite{Nelson}). As Makri and Miller
\cite{makmil1,makmil2} have shown, the speed of convergence of formula
(\ref{Trotter}) is far superior to that obtained by using the usual
Feynman-type approaches and is particularly well adapted to numerical
calculations using Monte-Carlo methods.

\subsection{A simple example}

Let $H$ be the Hamiltonian function of the one-dimensional harmonic oscillator%
\[
H=\frac{1}{2m}(p^{2}+m^{2}\omega^{2}x^{2});
\]
choosing $t_{0}=0$ the action integral $S$ is here given by
\[
S(x,x_{0},t)=\frac{1}{2\sin\omega t}((x^{2}+x_{0}^{2})\cos\omega t-2xx_{0})
\]
and the exact propagator for the corresponding Schr\"{o}dinger equation%
\[
i\hbar\frac{\partial\Psi}{\partial t}(x)=\left[  -\frac{\hbar^{2}}{2m}%
\frac{\partial^{2}}{\partial x^{2}}+\frac{m\omega^{2}}{2}x^{2}\right]  \Psi(x)
\]
is the Mehler kernel%
\begin{equation}
K(x,x_{0},t)=\left(  \frac{m\omega}{2\pi i\hbar\sin\omega t}\right)
^{1/2}\exp\left[  \frac{i}{2\hbar\sin\omega t}((x^{2}+x_{0}^{2})\cos\omega
t-2xx_{0})\right]  \label{mehler}%
\end{equation}
where the argument of the first factor in the left-hand side is chosen so that
$\lim_{t\rightarrow0}K(x,x_{0},t)=\delta(x-x_{0})$. Using second order Taylor
expansions of $\sin\omega t$ and $\cos\omega t$ at the origin one approximates
$S(x,x_{0},t)$ by
\[
\overline{S}(x,x_{0},t)=m\frac{(x-x_{0})^{2}}{2t}-\frac{m\omega^{2}}{2}%
(x^{2}+xx_{0}+x_{0}^{2})t
\]
which is precisely the value one obtains using formulas (\ref{sbar}) and
(\ref{vbar}) since we have%
\begin{align*}
\overline{V}(x,x_{0}) &  =\frac{m\omega^{2}}{2}\int_{0}^{1}(\tau
x+(1-\tau)x_{0})^{2}d\tau\\
&  =\frac{m\omega^{2}}{2}(x^{2}+xx_{0}+x_{0}^{2}).
\end{align*}
The corresponding Kerner--Sutcliffe propagator is thus given by%
\begin{equation}
\overline{K}(x,x_{0},t)=\left(  \frac{m}{2\pi i\hbar t}\right)  ^{1/2}%
\exp\left[  \frac{i}{\hbar}\left(  m\frac{(x-x_{0})^{2}}{2t}-\frac{m\omega
^{2}}{2}(x^{2}+xx_{0}+x_{0}^{2})\right)  \right]
\end{equation}
which is in agreement with formula (\ref{kbarfree}).

\section{Bohmian Trajectories\label{sec2}}

\subsection{Quantum motion}

The idea lying behind the Bohmian approach (Bohm and Hiley \cite{BoHi}, Hiley
\cite{Hiley1}, Hiley and collaborators \cite{hica,hicama}, Holland
\cite{Holland}, Wyatt \cite{Wyatt}) is the following: let $\Psi$ be the
wavefunction solution of Schr\"{o}dinger's equation (\ref{schrodeq}); we
assume we assume that $\Psi(x)$ is infinitely differentiable. Writing $\Psi$
in polar form $\sqrt{\rho}e^{iS/\hbar}$, this equation, as we have already
indicated, is equivalent, for $\rho(x,t)\neq0$, to the coupled system of
partial differential equations:
\begin{equation}
\frac{\partial S}{\partial t}+\sum_{j=1}^{n}\frac{1}{2m_{j}}\left(
\frac{\partial S}{\partial x_{j}}\right)  ^{2}+V+Q^{\Psi}=0\label{HJ1}%
\end{equation}
where the term
\begin{equation}
Q^{\Psi}=-\sum_{j=1}^{n}\frac{\hbar^{2}}{2m_{j}}\frac{1}{\sqrt{\rho}}%
\frac{\partial^{2}\sqrt{\rho}}{\partial x_{j}^{2}}\label{qp}%
\end{equation}
is Bohm's quantum potential; equivalently
\begin{equation}
Q^{\Psi}=-\sum_{j=1}^{n}\frac{\hbar^{2}}{4m_{j}}\left(  \frac{1}{\rho}%
\frac{\partial^{2}\rho}{\partial x_{j}^{2}}-\frac{1}{2\rho^{2}}\left(
\frac{\partial\rho}{\partial x_{j}}\right)  ^{2}\right)  .\label{qpbis}%
\end{equation}
Notice that the quantum potential is usually time-dependent: $Q^{\Psi}%
=Q^{\Psi}(x,t)$ since it is expressed in terms of the time-dependent function
$\rho(x,t)$. We will assume that $\rho_{0}(x)=\rho(x,t_{0})$ does not vanish,
that is $\rho_{0}(x)>0$ for all $x$ in some domain $\Omega$. It follows by
continuity that we will also have $\rho(x,t)>0$ provided that $\Delta
t=t-t_{0}$ is small enough.

Equation (\ref{HJ1}) is, mathematically speaking, just the Hamilton--Jacobi
equation (or Euler equation in fluid dynamics) for $H^{\Psi}=H+Q^{\Psi}$, and
\begin{equation}
\frac{\partial\rho}{\partial t}+\sum_{j=1}^{n}\frac{1}{m_{j}}\frac{\partial
}{\partial x_{j}}\left(  \rho\frac{\partial S}{\partial x_{j}}\right)  =0
\label{CO1}%
\end{equation}
is the equation of continuity familiar again from hydrodynamics \cite{fluid};
it ensures the conservation of probability (see the recent work by Heifetz and
Cohen \cite{heco16} for a detailed discussion of these equations in terms of
the Madelung fluid). In Bohm's theory of motion the trajectory of the system
of particles is determined by the equations
\begin{equation}
m_{j}\dot{x}_{j}^{\Psi}=\frac{\partial S}{\partial x_{j}}(x^{\Psi},t)\text{
\ , \ }x^{\Psi}(t_{0})=x_{0} \label{Bohm}%
\end{equation}
where $x_{0}=(x_{0,1},...,x_{0,n})$ is the initial position.

At this point we wish to make the following remark: the simple derivation of
equations (\ref{HJ1}) to (\ref{CO1}) obscures a deeper mathematical relation
between the Hilbert space formalism of quantum mechanics and the Hamiltonian
flows of classical mechanics. This exact relationship has been derived in very
general terms by de Gosson and Hiley \cite{mdgbh10}; for a detailed
mathematical study of that correspondence see the recent paper by de Gosson
\cite{go15}. There is a \emph{one-to-one} and \emph{onto} correspondence
between Hamiltonian flows generated by a Hamiltonian $H$ and the strongly
continuous unitary one-parameter evolution groups $(U_{t})$ satisfying
Schr\"{o}dinger's equation with the Hamiltonian operator $H(x,-i\hbar
\nabla_{x})$. This relation exploits the \emph{metaplectic} representation of
the underlying symplectic structure \cite{Birk,go15} of classical mechanics in
its Hamiltonian formulation. It is the metaplectic structure that gives rise
to the quantum properties. Since the classical and quantum motions are related
but different, it was proposed in de Gosson \cite{ICP} to call the object that
obeys the Bohmian law of motion (\ref{Bohm}) a \emph{metatron}.

\subsection{The Hamiltonian character of Bohmian trajectories}

\label{sec:hbt}

Let $p_{0}=(p_{1,0},...,p_{n,0})$ be an arbitrary momentum vector, and set%
\begin{equation}
p_{0}=-\nabla_{x_{0}}S(x,x_{0};t,t_{0}). \label{po}%
\end{equation}
Viewing $x,x_{0}$ as independent variables, the action $S(x,x_{0};t,t_{0})$ is
a generating function for $t-t_{0}\neq0$ sufficiently small
\cite{Arnold,HGoldstein,ICP,Birk}; in particular the Hessian matrix
\begin{equation}
S_{x,x_{0}}^{\prime\prime}=\left(  \frac{\partial^{2}S}{\partial x_{j}\partial
x_{0,k}}\right)  _{1\leq j,k\leq n} \label{hessian}%
\end{equation}
is invertible: $\det S_{x,x_{0}}^{\prime\prime}\neq0$. It follows from the
implicit function theorem that equation (\ref{po}) determines a function
$x=x^{\Psi}(t)$ (depending on $x_{0}$ and $t_{0}$ viewed as parameters),
defined by
\begin{equation}
p_{0}=-\nabla_{x_{0}}S(x^{\Psi}(t),x_{0};t,t_{0}). \label{pot}%
\end{equation}
Setting
\begin{equation}
p^{\Psi}(t)=\nabla_{x}S(x^{\Psi}(t),x_{0};t,t_{0}) \label{pt}%
\end{equation}
we claim that the functions $x^{\Psi}(t)$ and $p^{\Psi}(t)$ thus defined are
solutions of the Hamilton equations
\begin{equation}
\dot{x}^{\Psi}(t)=\nabla_{p}H^{\Psi}(x^{\Psi}(t),p^{\Psi}(t),t)\text{ ,
\ }\dot{p}^{\Psi}(t)=-\nabla_{x}H^{\Psi}(x^{\Psi}(t),p^{\Psi}(t),t))
\label{Hamilton}%
\end{equation}
with initial conditions $x^{\Psi}(t_{0})=x_{0}$, $p^{\Psi}(t_{0})=p_{0}$; here
$H^{\Psi}=H+Q^{\Psi}$ where $Q^{\Psi}$ is the quantum potential (\ref{qp}). We
are going to use classical Hamilton--Jacobi theory. For notational simplicity
we assume here that $n=1$; the generalization to the case of an arbitrary
number $n$ of degrees of freedom is straightforward. The quantum
Hamilton--Jacobi equation%
\begin{equation}
\frac{\partial S}{\partial t}+H^{\Psi}\left(  x,\frac{\partial S}{\partial
x},t\right)  =0 \label{HJ3}%
\end{equation}
can be treated mathematically using Hamilton--Jacobi theory. Differentiating
both sides of (\ref{HJ3}) with respect to $p$ and using the chain rule, we get%
\begin{equation}
\frac{\partial^{2}S}{\partial x_{0}\partial t}+\frac{\partial H^{\Psi}%
}{\partial p}\frac{\partial^{2}S}{\partial x_{0}\partial x}=0 \label{a}%
\end{equation}
and differentiating equation (\ref{pot}) with respect to time $t$ yields%
\begin{equation}
\frac{\partial^{2}S}{\partial x_{0}\partial t}+\frac{\partial^{2}S}{\partial
x\partial x_{0}}\dot{x}^{\Psi}=0. \label{b}%
\end{equation}
Subtracting (\ref{b}) from (\ref{a}) we get%
\[
\frac{\partial^{2}S}{\partial x\partial x_{0}}\left(  \frac{\partial H^{\Psi}%
}{\partial p}-\dot{x}^{\Psi}\right)  =0
\]
which produces the first Hamilton equation (\ref{Hamilton}) since it is
assumed that we have $\partial^{2}S/\partial x\partial x_{0}\neq0$ (condition
(\ref{hessian}) in the case $n=1$). Let us next show that the second Hamilton
equation (\ref{Hamilton}) is satisfied as well. For this we differentiate the
quantum Hamilton--Jacobi equation (\ref{HJ3}) with respect to $x$, which
yields%
\begin{equation}
\frac{\partial^{2}S}{\partial x\partial t}+\frac{\partial H^{\Psi}}{\partial
x}+\frac{\partial H^{\Psi}}{\partial p}\frac{\partial^{2}S}{\partial x^{2}}=0.
\label{c}%
\end{equation}
Differentiating the equality (\ref{pt}) with respect to $t$ we get%
\begin{equation}
\frac{\partial^{2}S}{\partial t\partial x}=-\dot{p}^{\Psi}(t)-\frac
{\partial^{2}S}{\partial x^{2}}\dot{x}^{\Psi} \label{d}%
\end{equation}
and hence equation (\ref{c}) can be rewritten as%
\[
-\dot{p}^{\Psi}-\frac{\partial^{2}S}{\partial x^{2}}\dot{x}^{\Psi}%
+\frac{\partial H^{\Psi}}{\partial x}+\frac{\partial H^{\Psi}}{\partial
p}\frac{\partial^{2}S}{\partial x^{2}}=0.
\]
Taking into account the relation $\dot{x}^{\Psi}=\partial H^{\Psi}/\partial p$
established above we have
\[
-\dot{p}^{\Psi}-\frac{\partial H^{\Psi}}{\partial x}=0
\]
which is precisely the second Hamilton equation (\ref{Hamilton}). There
remains to show that the initial value conditions $x^{\Psi}(t_{0})=x_{0}$ and
$p^{\Psi}(t_{0})=p_{0}$ are satisfied. Recalling that $x^{\Psi}(t)$ is defined
by the implicit equation
\[
p_{0}=-\frac{\partial S}{\partial x_{0}}(x^{\Psi}(t),x_{0};t,t_{0})
\]
(equation (\ref{pot})) the formulas (\ref{estimateICP}), (\ref{sbar}) imply
that%
\[
p_{0}=\frac{m(x^{\Psi}(t)-x_{0})}{t-t_{0}}-\frac{\partial\overline{V}%
}{\partial x_{0}}(x^{\Psi}(t),x_{0})(t-t_{0})+\mathcal{O}((t-t_{0})^{2});
\]
taking the limit $t\rightarrow t_{0}$ and noting that
\begin{equation}
\frac{\partial\overline{V}}{\partial x}(x_{0})=\int_{0}^{1}\tau\frac{\partial
V}{\partial x}(\tau x_{0}+(1-\tau)x_{0})d\tau=\frac{1}{2}\frac{\partial
V}{\partial x}(x_{0}) \label{limto}%
\end{equation}
we get both $x^{\Psi}(t_{0})=x_{0}$ and $p_{0}=m\dot{x}^{\Psi}(t_{0})=p^{\Psi
}(t_{0})$.

We refer to the papers by Holland \cite{Holland1,Holland2} for a thorough
discussion of the interpretation of Bohmian trajectories from the Hamiltonian
point of view.

To complete our discussion, we make the two following observations:

\begin{itemize}
\item Even when the Hamiltonian function $H$ does not depend explicitly on
time, the function $H^{\Psi}=H+Q^{\Psi}$ is usually time-dependent (because
the quantum potential generally is), so the flow $(f_{t}^{\Psi})$ it
determines does not inherit the usual group property $f_{t}f_{t^{\prime}%
}=f_{t+t^{\prime}}$ of the flow determined by the classical Hamiltonian $H$.
One has to use instead the \textquotedblleft time-dependent
flow\textquotedblright\ $(f_{t,t^{\prime}}^{\Psi})$, defined by
$f_{t,t^{\prime}}^{\Psi}=f_{t}^{\Psi}(f_{t^{\prime}}^{\Psi})^{-1}$, which has
a groupoid property \cite{AM,AMR} in the sense that $f_{t,t^{\prime}}^{\Psi
}f_{t^{\prime},t^{\prime\prime}}^{\Psi}=f_{t,t^{\prime\prime}}^{\Psi}$ (the
Chapman--Kolmogorov law, which expresses \emph{causality});

\item The time-dependent flow\ $(f_{t,t^{\prime}}^{\Psi})$ consists of
canonical transformations; that is, the Jacobian matrix of $f_{t,t^{\prime}%
}^{\Psi}$ calculated at any point $(x,p)$, where it is defined by a symplectic
matrix. This is an immediate consequence of the fact discussed above, namely,
that the flow determined by \emph{any} Hamiltonian function has this property
\cite{Arnold,HGoldstein,ICP,Birk}. There is thus a one-to-one correspondence
between the quantum flow $(f_{t}^{\Psi})$ and the classical Hamiltonian flow
$(f_{t})$ (see de Gosson \cite{go15} for a detailed study of this correspondence).
\end{itemize}

\subsection{Short-time solutions}

We have seen that the Bohmian trajectory for a particle initially sharply
localized at a point $x_{0}$ is Hamiltonian, and in fact governed by the
Hamilton equations (\ref{Hamilton}):%
\begin{equation}
\dot{x}=\nabla_{p}H^{\Psi}(x,p,t)\text{ \ , \ }\dot{p}=-\nabla_{x}H^{\Psi
}(x,p,t). \label{hameq}%
\end{equation}
corresponding to the $\Psi$-dependent Hamiltonian function $H^{\Psi}%
=H+Q^{\Psi}$. Hereinafter, a time interval $\Delta t=t-t_{0}$ shall be
considered small if for all $j$, $\Delta t \ll\Delta x_{j}(t)/v_{j}$, where
$\Delta x_{j}(t)=x_{j}^{\Psi}(t)-x_{j}(t_{0})$ and $v_{j}=p_{j}(t_{0})/m_{j}$.
We are going to show that for short time intervals $\Delta t$ these solutions
are, for given initial data $x_{0},~p_{0}$, identical to the solutions of the
usual Hamilton equations%
\[
\dot{x}=\nabla_{p}H(x,p,t)\text{ \ , \ }\dot{p}=-\nabla_{x}H(x,p,t)
\]
up to the order $\mathcal{O}(\Delta t^{2})$.

Because of the importance of this result we dignify it as a theorem:

\begin{theorem}
\label{Thm1}Let $x^{\Psi}(t)=(x_{1}^{\Psi}(t),...,x_{n}^{\Psi}(t))$ and
$p^{\Psi}(t)=(p_{1}^{\Psi}(t),...,p_{n}^{\Psi}(t))$ be the solution of
Hamilton's equations for $H^{\Psi}=H+Q^{\Psi}$. For $t$ close to the initial
time $t_{0}$ we have
\begin{align}
x_{j}^{\Psi}(t)  &  =x_{0,j}+\frac{p_{0,j}}{m_{j}}(t-t_{0})+\mathcal{O}%
((t-t_{0})^{2})\label{good1}\\
p_{j}^{\Psi}(t)  &  =p_{0,j}-\frac{1}{m_{j}}\frac{\partial V}{\partial x_{j}%
}(x_{0})(t-t_{0})+\mathcal{O}((t-t_{0})^{2}). \label{good2}%
\end{align}

\end{theorem}

\begin{proof}
We are going to prove these asymptotic formulas for $n=1$; the general case is
a straightforward generalization, working separately on each coordinate. We
notice that formula (\ref{good1}) is an immediate consequence of formula
(\ref{good2}) since
\[
\dot{x}_{j}^{\Psi}=\frac{\partial H^{\Psi}}{\partial p_{j}}=\frac{p_{j}}%
{m_{j}}.
\]
To prove formula (\ref{good2}) we begin showing that we have
\begin{equation}
m\dot{x}^{\Psi}=m\frac{x^{\Psi}-x_{0}}{t-t_{0}}-\frac{1}{2}\frac{\partial
V}{\partial x}(x_{0})(t-t_{0})+\mathcal{O}((t-t_{0})^{2})). \label{master}%
\end{equation}
Recalling that by the first formula (\ref{Bohm}) we have%
\begin{equation}
m\dot{x}^{\Psi}=\frac{\partial S}{\partial x}(x^{\Psi},t). \label{Bohmbis}%
\end{equation}
we replace $S$ with its approximation $\overline{S}$; taking
(\ref{estimateICP}) into account, this yields
\begin{equation}
m\dot{x}^{\Psi}=m\frac{x^{\Psi}-x_{0}}{t-t_{0}}-\frac{\partial\overline{V}%
}{\partial x}(x^{\Psi},x_{0})(t-t_{0})+\mathcal{O}((t-t_{0})^{2}). \label{eqn}%
\end{equation}
By continuity we have $x^{\Psi}=x_{0}+\mathcal{O}(t-t_{0})$, hence using
formula (\ref{limto}),
\begin{align*}
\frac{\partial\overline{V}}{\partial x}(x^{\Psi},x_{0})  &  =\frac
{\partial\overline{V}}{\partial x}(x_{0},x_{0})+\mathcal{O}(t-t_{0})\\
&  =\frac{1}{2}\frac{\partial V}{\partial x}(x_{0})+\mathcal{O}(t-t_{0});
\end{align*}
insertion of this value in (\ref{eqn}) yields (\ref{master}). We are now going
to use equation (\ref{master}) to derive formulas (\ref{good1}) and
(\ref{good2}). Differentiating both sides of (\ref{master}) with respect to
time $t$ we get%
\[
m\ddot{x}^{\Psi}=-m\frac{x^{\Psi}-x_{0}}{(t-t_{0})^{2}}+m\frac{\dot{x}^{\Psi}%
}{t-t_{0}}-\frac{1}{2}\frac{\partial V}{\partial x}(x_{0})+\mathcal{O}%
(t-t_{0});
\]
replacing in this equality $\dot{x}^{\Psi}$ with the value given by
(\ref{master}) and simplifying leads to%
\[
m\ddot{x}^{\Psi}=-\frac{\partial V}{\partial x}(x_{0})+\mathcal{O}(t-t_{0}).
\]
Integrating both sides from $t_{0}$ to $t$ yields%
\[
m\dot{x}^{\Psi}-m\dot{x}^{\Psi}(0)=-\frac{\partial V}{\partial x}%
(x_{0})(t-t_{0})+\mathcal{O}((t-t_{0})^{2})
\]
which is formula (\ref{good2}), since $p^{\Psi}=m\dot{x}^{\Psi}$.
\end{proof}

The quantum potential is absent from the asymptotic solutions (\ref{good1})
and (\ref{good2}); this suggests that the latter is negligible for short
times. Let us show that this is indeed the case. Recall that we are dealing
with a point-like source modeled by the Dirac delta centered at $x_{0}$; the
wavefunction is thus the unique solution to Schr\"{o}dinger's equation
(\ref{schrodeq}) with initial datum $\Psi(x,t_{0})=\delta(x-x_{0})$; in view
of formula (\ref{kernel}) expressing the solution in terms of the propagator
$K$ we thus have
\[
\Psi(x,t)=\int K(x,x^{\prime},t,t_{0})\delta(x^{\prime}-x_{0})d^{n}x^{\prime
}=K(x,x_{0},t,t_{0})
\]
hence the wavefunction is the propagator $K(x,x_{0},t,t_{0})$ itself. Writing
$\Psi=\sqrt{\rho}e^{\frac{i}{\hbar}S}$ the wavefunction is thus
\[
\Psi(x,\Delta t)=\sqrt{\rho(x,\Delta t)}e^{\frac{i}{\hbar}S(x,\Delta t)}%
\]
after time $\Delta t=t-t_{0}$. Taking into account the estimate
(\ref{essentiel}) for the approximate short-time propagator (\ref{kbarfree})%
\[
\overline{\Psi}(x,\Delta t)=\overline{K}(x,x_{0},t,t_{0})=\left(  \frac
{m_{1}\cdot\cdot\cdot m_{n}}{2\pi i\hbar(t-t_{0})}\right)  ^{n/2}e^{\frac
{i}{\hbar}\overline{S}(x,x_{0},t,t_{t})}%
\]
we have, using the inequality $|\;|a|-|b|\;|\leq|a-b|$ (which is a
straightforward consequence of the triangle inequality) with $a=|\Psi|$,
$b=|\overline{\Psi}|$,
\[
|\;|\Psi(x,\Delta t)|-|\overline{\Psi}(x,\Delta t)|\;|\leq|\Psi(x,\Delta
t)-\overline{\Psi}(x,\Delta t)|=\mathcal{O}(\Delta t^{2})
\]
and hence%
\begin{align*}
\sqrt{\rho(x,\Delta t)}  &  =\left(  \frac{m_{1}\cdot\cdot\cdot m_{n}}{2\pi
i\hbar\Delta t}\right)  ^{n/2}+\mathcal{O}(\Delta t^{2})\\
&  =\left(  \frac{m_{1}\cdot\cdot\cdot m_{n}}{2\pi i\hbar\Delta t}\right)
^{n/2}(1+\mathcal{O}(\Delta t^{\frac{n}{2}+2}))
\end{align*}
It immediately follows that%
\[
Q^{\Psi}(x,\Delta t)=\mathcal{O}(\Delta t^{\frac{n}{2}+2})
\]
(the quantum potential $Q^{\Psi}(x,t)$ is not defined for $t=0$ when $x\neq
x_{0}$ in the present case; the limit $\lim_{\Delta t\rightarrow0}Q^{\Psi
}(x,\Delta t)$ is thus singular). We notice that the quantum potential
decreases for small times as the number of degrees of freedom increases; the
effect of the quantum potential is thus negligible for systems consisting of a
large number of particles and vanishes at the macroscopic scale. As will be
further discussed in the last section, this might have further bearings on the
way we understand decoherence.

That the quantum evolution becomes classical for infinitesimal times has been
hinted by many authors; it was apparently known already to Feynman \cite{FH}
(see the discussion in \cite{Ali}); in \cite{Holland} (p. 269) Holland has
discussed this fact, by considering infinitesimal time intervals whose
sequence constructs a finite path. He shows that along each segment the motion
is classical due to a negligible quantum potential, and that it follows that
the quantum path may be decomposed into a sequence of segments along each of
which the classical action is a minimum.

In \cite{MaSch} Markowsky and Schopohl briefly discuss this in the context of
cold Bose atoms around the crossing of quantum waveguides.

\subsection{The general case\label{subsec2.1}}

In the previous discussion we assumed that the wavefunction was perfectly
localized at initial time: $\Psi(x,t_{0})=\delta(x-x_{0})$, in which case
$\Psi(x,t)$ is just the propagator $K(x,x_{0},t,t_{0})$. Let us now consider
the case where $\Psi(x,t_{0})$ is an arbitrary wavepacket $\Psi_{0}%
(x)=\sqrt{\rho_{0}(x)}e^{iS_{0}(x)/\hbar}$. Writing the solution of
Schr\"{o}dinger's equation (\ref{schrodeq}) as%
\[
\Psi(x,t)=\sqrt{\rho(x,t)}e^{\frac{i}{\hbar}S(x,t)}%
\]
the phase $S(x,t)$ satisfies the quantum Hamilton--Jacobi equation
\begin{equation}
\frac{\partial S}{\partial t}(x,t)+\sum_{j=1}^{n}\frac{1}{2m_{j}}\left(
\frac{\partial S}{\partial x_{j}}(x,t)\right)  ^{2}+V(x)+Q^{\Psi}(x,t)=0
\label{HJ1bis}%
\end{equation}
where the quantum potential is
\begin{equation}
Q^{\Psi}(x,t)=-\sum_{j=1}^{n}\frac{\hbar^{2}}{2m_{j}}\frac{1}{\sqrt{\rho
(x,t)}}\frac{\partial^{2}\sqrt{\rho(x,t)}}{\partial x_{j}^{2}}. \label{qpsi}%
\end{equation}
The continuity equation is here
\begin{equation}
\frac{\partial\rho(x,t)}{\partial t}+\sum_{j=1}^{n}\frac{1}{m_{j}}%
\frac{\partial}{\partial x_{j}}\left(  \rho(x,t)\frac{\partial S(x,t)}%
{\partial x_{j}}\right)  =0. \label{CO2}%
\end{equation}
In terms of $\rho$, a time interval $\Delta t$ may be considered small if
$\Delta t = t-t_{0} \ll\rho(x,t)/\frac{ \partial\rho(x,t)}{\partial t}$ for
all $x$.

The Bohmian trajectory of the system of particles is determined by the
equations
\begin{equation}
m_{j}\dot{x}_{j}^{\Psi}=\frac{\partial S}{\partial x_{j}}(x^{\Psi},t)\text{
\ , \ }x_{j}^{\Psi}(t_{0})=x_{0,j}. \label{Bohm2}%
\end{equation}
One shows that again this motion is given in phase space by the quantum
Hamilton equations%
\begin{equation}
\dot{x}_{j}^{\Psi}(t)=\frac{\partial H^{\Psi}}{\partial p_{j}}(x_{j}^{\Psi
}(t),p_{j}^{\Psi}(t),t)\text{ \ , \ }\dot{p}_{j}^{\Psi}(t)=-\frac{\partial
H^{\Psi}}{\partial x_{j}}(x_{j}^{\Psi}(t),p_{j}^{\Psi}(t),t) \label{Hamilton2}%
\end{equation}
with initial conditions
\[
x_{j}^{\Psi}(t_{0})=x_{0,j}\text{ \ , \ }p_{j}^{\Psi}(t_{0})=p_{0,j}%
=\frac{\partial S}{\partial x_{j}}(x_{0},t_{0});
\]
as before $H^{\Psi}=H+Q^{\Psi}$.

Let us now prove what can perhaps be seen as the main property of the flow
determined by equations (\ref{Hamilton2}), namely that for short times the
quantum potential is conserved along the trajectory up to an error term
$\mathcal{O}((t-t_{0})^{2})$. The analogue of Theorem \ref{Thm1} for the
general case will follow.

\begin{theorem}
\label{Thm2}Let $(x_{0},t_{0})$ be a fixed initial point of space-time. The
quantum potential $Q^{\Psi}$, calculated along the Bohmian trajectory
$x^{\Psi}(t)$ satisfies%
\begin{equation}
Q^{\Psi}(x^{\Psi}(t),t)=Q^{\Psi}(x_{0},t_{0})+\mathcal{O}((t-t_{0})^{2})
\label{smallq}%
\end{equation}
for small $\Delta t=t-t_{0}$.
\end{theorem}

\begin{proof}
Writing $\rho^{1/2}(x,t)=\sqrt{\rho(x,t)}$, $x^{\Psi}(t)=x^{\Psi}$ the quantum
potential at $(x_{0},t_{0})$ is
\[
Q^{\Psi}(x_{0},t_{0})=-\sum_{j=1}^{n}\frac{\hbar^{2}}{2m_{j}}\frac{1}%
{\rho^{1/2}(x_{0},t_{0})}\frac{\partial^{2}\rho^{1/2}}{\partial x_{j}^{2}%
}(x_{0},t_{0})
\]
and at $(x^{\Psi},t)$
\[
Q^{\Psi}(x^{\Psi},t)=-\sum_{j=1}^{n}\frac{\hbar^{2}}{2m_{j}}\frac{1}%
{\rho^{1/2}(x^{\Psi},t)}\frac{\partial^{2}\rho^{1/2}}{\partial x_{j}^{2}%
}(x^{\Psi},t).
\]
Defining the velocity field $v(x,t)=(v_{1}(x,t),...,v_{n}(x,t))$ by
\[
m_{j}v_{j}(x,t)=\frac{\partial S(x,t)}{\partial x_{j}}%
\]
the continuity equation (\ref{CO2}) can be written in the usual form%
\begin{equation}
\frac{\partial\rho(x,t)}{\partial t}+\nabla_{x}\cdot\left(  \rho
(x,t)v(x,t)\right)  =0.
\end{equation}
It follows that (Chorin and Marsden \cite{fluid}, \S 1.1)
\begin{equation}
\rho(x^{\Psi},t)=\rho(x_{0},t_{0})\left\vert \frac{dx^{\Psi}}{dx_{0}%
}\right\vert ^{-1} \label{fluid1}%
\end{equation}
where $dx^{\Psi}/dx_{0}$ is the determinant of the Jacobian matrix
\[
\frac{dx^{\Psi}}{dx_{0}}=\left(  \frac{\partial x_{j}^{\Psi}}{\partial
x_{0,k}}\right)  _{1\leq j,k\leq n}%
\]
of the flow map $x_{0}\longmapsto x^{\Psi}$. Now,%
\[
x_{j}^{\Psi}(t)=x_{0,j}+v_{0,j}\Delta t+\mathcal{O}(\Delta t^{2})
\]
hence%
\begin{equation}
\frac{\partial x_{j}^{\Psi}(t)}{\partial x_{0,k}}=\delta_{jk}+\mathcal{O}%
(\Delta t^{2}) \label{deltajk}%
\end{equation}
so that
\[
\left\vert \frac{dx^{\Psi}}{dx_{0}}\right\vert =\det(I+\mathcal{O}(\Delta
t^{2}))=1+\mathcal{O}(\Delta t^{2})
\]
then (\ref{fluid1}) yields%
\begin{equation}
\rho(x^{\Psi},t)=\rho(x_{0},t_{0})(1+\mathcal{O}(\Delta t^{2}))
\label{rhodelta}%
\end{equation}
and therefore also%
\begin{equation}
\rho^{1/2}(x^{\Psi},t)=\rho^{1/2}(x_{0},t_{0})(1+\mathcal{O}(\Delta t^{2})).
\label{rhohalf}%
\end{equation}
Differentiating this expression with respect to the initial variables
$x_{0,j}$ we get, using the chain rule together with (\ref{deltajk}),%
\begin{align*}
\frac{\partial}{\partial x_{0,j}}\left[  \rho^{1/2}(x^{\Psi},t)\right]   &
=\sum_{k=1}^{n}\frac{\partial\rho^{1/2}}{\partial x_{0,k}}(x^{\Psi}%
,t)\frac{\partial x_{k}^{\Psi}}{\partial x_{0,j}}\\
&  =\frac{\partial\rho^{1/2}}{\partial x_{0,j}}(x^{\Psi},t)(1+\mathcal{O}%
(\Delta t^{2}));
\end{align*}
repeating the same argument with $\rho^{1/2}$ replaced with $\partial
\rho^{1/2}/\partial x_{0,j}$ we get%
\[
\frac{\partial^{2}}{\partial x_{0,j}^{2}}\left[  \rho^{1/2}(x^{\Psi
},t)\right]  =\frac{\partial^{2}\rho^{1/2}}{\partial x_{0,j}^{2}}(x^{\Psi
},t)(1+\mathcal{O}(\Delta t^{2}))
\]
and hence
\[
\frac{\partial^{2}\rho^{1/2}}{\partial x_{0,j}^{2}}(x^{\Psi},t)=\frac
{\partial^{2}}{\partial x_{0,j}^{2}}\left[  \rho^{1/2}(x^{\Psi},t)\right]
(1+\mathcal{O}(\Delta t^{2}));
\]
in view of (\ref{rhohalf}) we thus have%
\[
\frac{\partial^{2}\rho^{1/2}}{\partial x_{0,j}^{2}}(x^{\Psi},t)=\frac
{\partial^{2}\rho^{1/2}}{\partial x_{0,j}^{2}}(x_{0},t_{0})(1+\mathcal{O}%
(\Delta t^{2})).
\]
Formula (\ref{smallq}) follows.
\end{proof}

A practical consequence of this result, might be the short time simulation of
various external potentials by shaping the initial wave packet to create the
suitable quantum potential. If one prepares the initial wavefunction with a
shape corresponding to a specific quantum potential, then according to the
above theorem, this potential might remain a good approximation for the actual
one along some short time interval.

We note that the crucial step in the proof above is formula (\ref{rhodelta}),
which follows from property (\ref{fluid1}) of the solution of the continuity
equation; all the other estimates we needed followed from it. Here is another
independent way to prove (\ref{rhodelta}); we limit ourselves to the case
$n=1$, in which we have%
\begin{align*}
\frac{d}{dt}\rho(x^{\Psi},t)  &  =\frac{\partial\rho}{\partial t}(x^{\Psi
},t)+\frac{\partial\rho}{\partial x}(x^{\Psi},t)\dot{x}^{\Psi}\\
&  =-\frac{\partial}{\partial x}(\rho(x^{\Psi},t)\dot{x}^{\Psi})+\frac
{\partial\rho}{\partial x}(x^{\Psi},t)\dot{x}^{\Psi}\\
&  =\rho(x^{\Psi},t)\frac{\partial\dot{x}^{\Psi}}{\partial x}.
\end{align*}
We have $x^{\Psi}=x_{0}+v_{0}t+\mathcal{O}(\Delta t^{2})$, hence $\dot
{x}^{\Psi}=v_{0}+\mathcal{O}(\Delta t)$ and $\partial\dot{x}^{\Psi}/\partial
x=\mathcal{O}(\Delta t)$, from which follows that%
\[
\frac{d}{dt}\rho(x^{\Psi},t)=\mathcal{O}(\Delta t);
\]
formula (\ref{rhodelta}) follows.

As observed in \cite{heco16} and \cite{MMadelung}, this suggests for short
times the incompressibility of the quantum fluid, which implies non-spreading
of the wavepacket and a constant Shannon entropy.

We now prove the following generalization of Theorem \ref{Thm1} for the short
time solutions of (\ref{Hamilton2}):

\begin{theorem}
\label{Thm3}Let $x^{\Psi}(t)=(x_{1}^{\Psi}(t),...,x_{n}^{\Psi}(t))$ and
$p^{\Psi}(t)=(p_{1}^{\Psi}(t),...,p_{n}^{\Psi}(t))$ be the solution at time
$t$ of Hamilton's equations for $H^{\Psi}=H+Q^{\Psi}$ where $Q^{\Psi} $ is
given by (\ref{qpsi}). (i) We have
\begin{align}
x_{j}^{\Psi}(t)  &  =x_{0,j}+\frac{p_{0,j}}{m_{j}}\Delta t+\mathcal{O}(\Delta
t^{2})\label{good3}\\
p_{j}^{\Psi}(t)  &  =p_{0,j}-\frac{\partial V}{\partial x_{j}}(x_{0})\Delta
t+\mathcal{O}(\Delta t^{2}); \label{good4}%
\end{align}
(ii) The quantum flow $(f_{t,t_{0}}^{\Psi})$ is approximated to order
$\mathcal{O}(\Delta t^{2})$ by the usual Hamiltonian flow $(f_{t,t_{0}})$:%
\begin{equation}
f_{t,t_{0}}^{\Psi}(x_{0},p_{0})=f_{t,t_{0}}(x_{0},p_{0})+\mathcal{O}(\Delta
t^{2}) \label{good5}%
\end{equation}

\end{theorem}

\begin{proof}
We again limit ourselves to the case $n=1$. As in the proof of Theorem
\ref{Thm1}, formula (\ref{good3}) immediately follows from formula
(\ref{good4}). The latter is a consequence of (\ref{smallq}), since we then
have, using the explicit expression of $H^{\Psi}$,
\begin{align*}
\dot{p}^{\Psi}(t)  &  =-\frac{\partial V}{\partial x}(x^{\Psi}(t))-\frac
{\partial Q}{\partial x}(x^{\Psi},t)\\
&  =-\frac{\partial V}{\partial x}(x^{\Psi}(t))-\frac{\partial}{\partial
x}Q(x_{0},t_{0})+\mathcal{O}(\Delta t^{2})\\
&  =-\frac{\partial V}{\partial x}(x^{\Psi}(t))+\mathcal{O}(\Delta t^{2})
\end{align*}
because $Q(x_{0},t_{0})$ is constant. Formula (\ref{good5}) follows since the
phase space transformations $g_{t,t_{0}}$ defined by $g_{t,t_{0}}(x_{0}%
,p_{0})=(x(t,t_{0}),p(t,t_{0}))$ with
\[
x_{j}(t,t_{0})=x_{0,j}+\frac{p_{0,j}}{m_{j}}\Delta t\text{ \ , \ }%
p_{j}(t,t_{0})=p_{0,j}-\frac{\partial V}{\partial x_{j}}(x_{0})\Delta t
\]
are themselves an $\mathcal{O}(\Delta t^{2})$ approximation to the
time-dependent Hamiltonian flow $(f_{t,t_{0}})$ in view of Taylor's theorem.
\end{proof}

\section{Quantum Zeno Effect\label{sec3}}

The quantum Zeno effect is widely discussed in the literature
\cite{ebpb91,widhjbdw,ej84,kk81}. It has been claimed by Misra and Sudarshan
\cite{bmes77, ccesbm77} that the quantum Zeno effect leads to the conclusion
that, and we quote, \textquotedblleft an unstable particle observed
continuously whether it has decayed or not will \emph{never} be found to
decay\textquotedblright. These authors confine their discussion specifically
to the case of $\alpha$-decay in a cloud or bubble chamber. Peres~\cite{ap80}
has considered a more general question, namely, \textquotedblleft If an
unstable quantum system is kept under continuous observation, will it
decay?\textquotedblright\ We will see how the question raised by Misra and
Sudarshan \cite{bmes77} in their discussion on $\alpha$-decay can be answered
using the Bohm--Hiley approach \cite{BoHi}. In order to inhibit the occurrence
of $\alpha$-decay, we must use a process that directly interacts with the
unstable nuclei. It is not sufficient to surround the nuclei with a passive
detection device like a bubble or cloud chamber. Merely detecting the $\alpha
$-particle outside the nucleus is not sufficient to inhibit the decay.

In de Gosson and Hiley \cite{gohi2,gohi3} it was assumed that the initial
wavepacket $\Psi_{0}$ was a point source modelled by the propagator itself;
here we consider the more realistic situation where $\Psi_{0}$ is an arbitrary
wavepacket. For this we will apply the machinery developed in Section
\ref{subsec2.1} together with a generalization of the Lie--Trotter formula.

\subsection{The Lie--Trotter formula for time-dependent flows}

We are going to prove a rather straightforward extension of the usual
Lie--Trotter \cite{Trotter} formula for flows to the time-dependent case; we
will use this result in the next section as we deal with the quantum Zeno
effect. Recall (Abraham \emph{et al.} \cite{AMR}, Chorin \emph{et al.}
\cite{chorinetal} and also Appendix B in de Gosson \cite{ICP}) that given a
vector field $X=X(x,p)=(a(x,p),b(x,p))$ on an open subset $\Omega$ of
$\mathbb{R}_{x,p}^{2n}$ (or any other Euclidean space, for that matter) a
family $(g_{t})$ of functions $\Omega\longrightarrow\mathbb{R}^{2n}$ is called
an \emph{algorithm} for the flow $(f_{t})$ determined by the vector field $X$
if for every point $z_{0}=(x_{0},p_{0})$ in $\Omega$ we have
\begin{equation}
X(z_{0})=\left.  \frac{\partial}{\partial t}k_{t}(z_{0})\right\vert _{t=0}.
\label{xz}%
\end{equation}
In this case the sequence of iterates $(k_{t/N}(z_{0}))^{N}$ converges towards
$f_{t}(z_{0})$:
\begin{equation}
f_{t}(z_{0})=\lim_{N\rightarrow\infty}(k_{t/N}(z_{0}))^{N}. \label{lt1}%
\end{equation}

Applying this result to the case where the vector field $X$ is generated by a
time-independent Hamiltonian $H$, that is $X=(\nabla_{p}H,-\nabla_{x}H)$,
formula (\ref{lt1}) yields a convenient procedure for approximating
Hamiltonian flows. In our context the trouble is that we would like to apply
the same kind of procedure to the Hamiltonian function $H^{\Psi}=H+Q^{\Psi}$,
which is time-dependent because of the presence of the quantum potential
$Q^{\Psi}$; as we have seen, the flow determined by such a Hamiltonian
function is no longer a one-parameter group (this is due to the fact that the
Hamiltonian vector field is here time-dependent: $X_{t}=X(x,p,t)$ and thus it
is not a vector field in the usual sense. The way out of this difficulty
consists of lifting the flow to the time-dependent phase space $\mathbb{R}%
_{x,p,t}^{2n+1}$: defining the \textquotedblleft suspended vector
field\textquotedblright\ $\widetilde{X}=(X,1)$ (Abraham and Marsden \cite{AM})
the flow $(\widetilde{f}_{t})$ it determines is given by the formula
\begin{equation}
\widetilde{f}_{t}(z_{0},t_{0})=(f_{t+t_{0},t_{0}}(z_{0}),t+t_{0})
\label{extended}%
\end{equation}
where, as usual $f_{t,t^{\prime}}=f_{t}(f_{t^{\prime}})^{-1}$ is the
time-dependent flow of $X_{t}$. On calls $(\widetilde{f}_{t})$ the suspended
flow of $X_{t}$. Now,%
\begin{align*}
\widetilde{f}_{t}\widetilde{f}_{t^{\prime}}(x_{0},p_{0},t_{0})  &
=\widetilde{f}_{t}(f_{t^{\prime}+t_{0},t_{0}}(z_{0}),t^{\prime}+t_{0})\\
&  =(f_{t+t^{\prime}+t_{0},t^{\prime}+t_{0}}f_{t^{\prime}+t_{0},t_{0}}%
(z_{0}),t+t^{\prime}+t_{0})\\
&  =(f_{t+t^{\prime}+t_{0},t_{0}}(z_{0}),t+t^{\prime}+t_{0})\\
&  =\widetilde{f}_{t+t^{\prime}}(z_{0},t_{0})
\end{align*}
hence $(\widetilde{f}_{t})$ is a one-parameter group, that is a flow in the
usual sense, to which we may apply the Lie--Trotter formula.

\begin{theorem}
\label{Thm4}Let $X_{t}=(a(x,p,t),b(x,p,t))$ be a time dependent vector field
on phase space $\mathbb{R}_{x,p}^{2n}$ and $(f_{t,t_{0}})$ its time-dependent
flow. Let $(k_{t,t_{0}})$ be a two-parameter family of transformations
$\mathbb{R}_{x,p}^{2n}\longrightarrow\mathbb{R}_{x,p}^{2n}$. If
\begin{equation}
X_{t_{0}}(z_{0})=\left.  \frac{\partial}{\partial t}k_{t,t_{0}}(z_{0}%
)\right\vert _{t=t_{0}}%
\end{equation}
then we have
\begin{equation}
\lim_{N\rightarrow\infty}(k_{t+t_{0},t_{N-1}+t_{0}}k_{t_{N-1}+t_{0}%
,t_{N-2}+t_{0}}\cdot\cdot\cdot k_{t_{1},t_{0}}(z_{0}))=f_{t,t_{0}}(z_{0})
\label{lt2}%
\end{equation}
for every $z_{0}=(x_{0},p_{0})$, where $t_{1},t_{2},...,t_{N-1}$ is a
subdivision of the interval $[t_{0},t]$ such that $|t_{j+1}-t_{j}%
|=|t-t_{0}|/N$.
\end{theorem}

\begin{proof}
Let $\widetilde{k}_{t}:\mathbb{R}_{x,p,t}^{2n+1}\longrightarrow\mathbb{R}%
_{x,p,t}^{2n+1}$ be defined by%
\begin{equation}
\widetilde{k}_{t}(z_{0},t_{0})=(k_{t+t_{0},t_{0}}(z_{0}),t+t_{0});
\label{ktilde}%
\end{equation}
we have
\begin{align*}
\left.  \frac{\partial}{\partial t}\widetilde{k}_{t}(z_{0},t_{0})\right\vert
_{t=t_{0}}  &  =\left(  \left.  \frac{\partial}{\partial t}k_{t+t_{0},t_{0}%
}(z_{0})\right\vert _{t=t_{0}},t_{0}\right) \\
&  =(X_{t_{0}}(z_{0}),t_{0})\\
&  =\widetilde{X}(z_{0},t_{0})
\end{align*}
hence $(\widetilde{k}_{t})$ is an algorithm for the suspended flow
$(\widetilde{f}_{t})$ of $X_{t}$. Applying the conventional Lie--Trotter
formula (\ref{lt1}) to the algorithm $(\widetilde{k}_{t})$ we have%
\[
\widetilde{f}_{t}(z_{0},t_{0})=\lim_{N\rightarrow\infty}(\widetilde{k}%
_{t/N})^{N}(z_{0},t_{0});
\]
Using formulas (\ref{ktilde}) and (\ref{extended}) one easily shows by
induction on the integer $N$ that
\begin{align*}
(\widetilde{k}_{t/N})^{N}(z_{0},t_{0})  &  =(k_{t+t_{0},\frac{(N-1)t}{N}%
+t_{0}}k_{\frac{(N-1)t}{N}+t_{0},\frac{(N-2)t}{N}+t_{0}}\cdot\cdot\cdot
k_{\frac{t}{N}+t_{0},t_{0}}(z_{0}),t+t_{0})\\
&  =k_{t+t_{0},t_{N-1}+t_{0}}k_{t_{N-1}+t_{0},t_{N-2}+t_{0}}\cdot\cdot\cdot
k_{t_{1},t_{0}}(z_{0})
\end{align*}
hence
\[
(f_{t+t_{0},t_{0}}(z_{0}),t+t_{0})=\lim_{N\rightarrow\infty}(k_{t,t_{N-1}%
}\cdot\cdot\cdot k_{t_{2},t_{1}}k_{t_{1},t_{0}}(z_{0}))
\]
which proves the generalized Lie--Trotter formula (\ref{lt2}).
\end{proof}

\subsection{The main result}

We now choose an initial time $t_{0}$ and denote the corresponding quantum
potential by $Q^{0}$:%
\begin{equation}
Q^{0}(x,t)=-\sum_{j=1}^{n}\frac{\hbar^{2}}{2m_{j}}\frac{1}{\sqrt{\rho
(x,t_{0})}}\frac{\partial^{2}\sqrt{\rho(x,t_{0})}}{\partial x_{j}^{2}}.
\label{qpsi0}%
\end{equation}
We set $H^{0}=H+Q^{0}$. After time $\Delta t=(t-t_{0})/N$ the particle is
observed at a point $x_{1}$, and the new wavefunction is now $\Psi_{1}%
=\sqrt{\rho_{1}}e^{iS_{1}/\hbar}$ hence the future quantum evolution of this
particle will be governed by the new Hamiltonian $H^{1}=H+Q^{1}$, where%
\begin{equation}
Q^{1}(x,t)=-\sum_{j=1}^{n}\frac{\hbar^{2}}{2m_{j}}\frac{1}{\sqrt{\rho
_{1}(x,t)}}\frac{\partial^{2}\sqrt{\rho_{1}(x,t)}}{\partial x_{j}^{2}}%
\end{equation}
together with the guiding condition $p_{1}=\nabla_{x}S_{1}$. Repeating this
procedure until time $t$ we obtain a sequence of points $x_{0},x_{1}%
,x_{2},...,x_{N}=x$ corresponding to the successive observations at times
$t_{0}$, $t_{1}=t_{0}+\Delta t$, $t_{2}=t_{0}+2\Delta t$,..., $t$. Let
$H^{0},H^{1},...,H^{N}$ be the sequence of Hamiltonian functions determined by
the quantum potentials $Q^{0},Q^{1},...,Q^{N-1}$:%
\[
H^{0}=H+Q^{0}\text{, }H^{1}=H+Q^{1}\text{,...,\ }H^{N-1}=H+Q^{N-1};
\]
we denote by $(f_{t,t_{0}}^{0})$, $(f_{t,t_{1}}^{1})$,...,$(f_{t,t_{N-1}%
}^{N-1})$ the corresponding time-dependent flows. Writing $z_{0}=(x_{0}%
,p_{0})$, $z_{1}=(x_{1},p_{1})$, etc. we thus have a sequence of successive
equalities%
\[
z_{1}=f_{t_{1},t_{0}}^{0}(z_{0})\text{, }z_{2}=f_{t_{2},t_{1}}^{1}%
(z_{1})\text{,..., }z_{N}=f_{t,t_{N-1}}^{N-1}(z_{N-1})
\]
where $p_{1}=\nabla_{x}S_{1}$, $p_{2}=\nabla_{x}S_{2}$,...,$p_{N-1}=\nabla
_{x}S_{N-1}$, which implies that the final position $x=x_{N}$ of the particle
at time $t$ is expressed in terms of the initial point $x_{0}$ by the formula%
\[
(x,p)=f_{t,t_{N-1}}^{N-1}\cdot\cdot\cdot f_{t_{2},t_{1}}^{1}f_{t_{1},t_{0}%
}^{0}(x_{0},p_{0})
\]
(notice that the intermediate points $t_{1},t_{2},...,t_{N-1}$ depend on $t$).

\begin{theorem}
\label{Thm5}Let the phase space transformations $f_{t_{1},t_{0}}^{0}%
,f_{t_{2},t_{1}}^{1}...,f_{t,t_{N-1}}^{N-1}$ be defined as above, $t_{1}%
,t_{2},...,t_{N-1}$ being a subdivision of the interval $[t_{0},t]$ such that
$\Delta t=t_{j+1}-t_{j}=(t-t_{0})/N$. The two-parameter family of
transformations $k_{t,t_{0}}$ defined by
\[
k_{t,t_{0}}(z_{0})=f_{t,t_{N-1}}^{N-1}\cdot\cdot\cdot f_{t_{2},t_{1}}%
^{1}f_{t_{1},t_{0}}^{0}(z_{0})
\]
is an algorithm for the flow $(f_{t})$ determined by the classical Hamiltonian
$H$.
\end{theorem}

\begin{proof}
It is sufficient to show that for small $\Delta t=t-t_{0}$ we have%
\begin{equation}
k_{t,t_{0}}(z_{0})=f_{t,t_{0}}(z_{0})+\mathcal{O}(\Delta t^{2}) \label{kf}%
\end{equation}
for then
\begin{align*}
\left.  \frac{\partial}{\partial t}k_{t,t_{0}}(z_{0})\right\vert _{t=t_{0}}
&  =\lim_{\Delta t\rightarrow0}\frac{k_{t_{0}+\Delta t,t_{0}}(z_{0})-z_{0}%
}{\Delta t}\\
&  =\lim_{\Delta t\rightarrow0}\frac{f_{t_{0}+\Delta t,t_{0}}(z_{0})-z_{0}%
}{\Delta t}\\
&  =X_{t}(z_{0})
\end{align*}
and $(k_{t,t_{0}})$ is then indeed an algorithm for $(f_{t,t_{0}})$ in view of
Theorem \ref{Thm4}. In light of Theorem \ref{Thm3} (formula (\ref{good5})),
each transformation $f_{t_{j+1},t_{j}}^{j}(z)$ is approximated to order
$\mathcal{O}((t_{j+1}-t_{j})^{2})=\frac{1}{N}\mathcal{O}(\Delta t^{2})$ by
$f_{t_{j+1},t_{j}}(z)$, hence
\begin{align*}
k_{t,t_{0}}(z_{0})  &  =f_{t,t_{N-1}}^{N-1}\cdot\cdot\cdot f_{t_{2},t_{1}}%
^{1}f_{t_{1},t_{0}}^{0}(z_{0})\\
&  =f_{t,t_{N-1}}\cdot\cdot\cdot f_{t_{2},t_{1}}f_{t_{1},t_{0}}(z_{0}%
)+\mathcal{O}(\Delta t^{2})
\end{align*}
which proves the estimate (\ref{kf}).
\end{proof}

We have shown that
\[
\lim_{N\rightarrow\infty}f_{t,t_{N-1}}^{N-1}\cdot\cdot\cdot f_{t_{2},t_{1}%
}^{1}f_{t_{1},0}^{0}(x_{0},p_{0})=f_{t}(x_{0},p_{0});
\]
the physical interpretation is that if one performs a series of observations
of the particle at very short time intervals, then the recorded trajectory is
the classical one. This is in agreement with the usual interpretation of the
quantum Zeno effect (\textquotedblleft the watched pot never
boils\textquotedblright): an almost continuous observation of a particle (or
system of particles) precludes the development of a quantum trajectory.

\subsection{The Mott problem}

The Mott problem (or paradox) illustrates in a striking manner the
difficulties of understanding the nature of wavepacket measurement. The story
goes back to 1929, when Heisenberg and Mott \cite{Mott,Hbg} reflected on the
following problem: in a Wilson cloud chamber, $\alpha$-particles are emitted
from nuclei of radioactive atoms; due to radial symmetry of the problem, the
$\alpha$-particles are represented by spherical wavefunctions. These particles
create droplets of condensation when interacting with water molecules in the
vapor, thus creating a track of condensation (see also Dell'Antonio
\cite{dell}). Since the spherical wavefunction spreads isotropically in all
directions it could have been expected to randomly ionize the water molecules,
leaving behind a spherical track. However, when one actually performs the
experiment one always finds a single linear track. In \cite{Bell}, Bell
illustrates the situation by comparing the cloud chamber with a stack of
photographic plates: to produce this straight trajectory, the ionized gas
molecules are assumed to act as an array of potential measuring devices,
leaving a record of the track of the $\alpha$-particle on the stack of plates.
We can thus regard the $\alpha$-particle as being \textquotedblleft
continuously watched\textquotedblright. This means, in a sense, that
continuous observation \textquotedblleft dequantizes\textquotedblright%
\ quantum trajectories. The frequent observations render the dynamic
contribution of the quantum potential negligible. This property is, of course,
essentially a consequence of the quantum Zeno effect, which has been shown to
inhibit the decay of unstable quantum systems when under continuous
observation (see \cite{BoHi, fapa09,gux,Hannabuss}). Another illustration of
how continuous observations of a different kind can give rise to a quantum
Zeno effect has already been given in Bohm and Hiley \cite{BoHi}. They
considered the transition of an Auger-like particle and showed that the
perturbed wavefunction, which is proportional to $\Delta t$ for times less
that $1/\Delta E$, ($\Delta E$ is the energy released in the transition) will
never become large and therefore cannot make a significant contribution to the
quantum potential necessary for the transition to occur. Thus again, the
reason that no transition will take place is the vanishing of the quantum potential.

\section{Discussion}

We have shown that the quantum potential plays a key role in determining the
behaviour of a quantum particle. This potential is not a mere
\textquotedblleft add-on\textquotedblright\ but an essential feature that is a
necessary consequence of the relationship between the symplectic group
structure (classical mechanics) and its double cover, the metaplectic group
and its non-linear generalisation. Specifically there is a deep relation
between the Hamilton flows, $f_{tt^{\prime}}$, of classical physics and the
flows $f_{tt^{\prime}}^{\psi}$ associated with the quantum behaviour in the
covering space~\cite{go15,mdgbh10}. It is this feature that is key to
understanding the relationship between a particle that exhibits only classical
behaviour and one that gives rise to quantum phenomena.

As is well known at the particle level, we have lost causality in quantum
phenomena but it has been replaced by what we will call \textquotedblleft wave
causality\textquotedblright, which is defined through the Schr\"{o}dinger
equation. However the real part of this equation under polar decomposition
appears to have restored causality at the particle level. How can this be?

The answer lies in the two observations we make in section \ref{sec:hbt} about
the relationship between the two types of flow, $f_{t,t^{\prime}}$ and
$f^{\psi}_{tt^{\prime}}$. This approach gives rise to a very different way of
analysing quantum phenomena from the conventional approaches, including the
approach of Bohmian mechanics as detailed in D\"{u}rr and Teufel~\cite{ddst09}%
.
Our two observations referred to the extensive work of one of us
(MdG~\cite{ICP}), who has shown in very general terms that causality as
defined by Chapman--Kolmogorov is retained in the covering space by the
appearance of an additional term, $Q^{\psi}$. Note that the Schr\"{o}dinger
equation operates in this covering space~\cite{vgss84}. Thus it is a necessary
feature in the relationship between classical and quantum mechanics. Having
established this relationship, it is now clear that once $Q^{\psi}$ becomes
small in relation to the kinetic energy, the phenomena loses it quantum
signature and becomes classical.

In this context we have analysed the short time behaviour of the quantum
potential and shown that the suppression of $Q^{\psi}$ is possible if the
successive positions of a particle can be defined in a short enough time. To
see this, we must examine equations (\ref{good3})--(\ref{good4}), that is
\begin{align}
x_{j}^{\Psi}(t)  &  =x_{0,j}+\frac{p_{0,j}}{m_{j}}\Delta t+\mathcal{O}(\Delta
t^{2})\\
p_{j}^{\Psi}(t)  &  =p_{0,j}-\frac{\partial V}{\partial x_{j}}(x_{0})\Delta
t+\mathcal{O}(\Delta t^{2}).
\end{align}
Notice that there is no quantum potential present in either equation. Only
when we allow higher order terms does the quantum potential appear. Thus if it
is possible to obtain information of succession of positions in a short enough
time without deflecting the particle significantly, then equation (\ref{lt2})
shows that no quantum potential will appear and the trajectory will be a
classical trajectory. In other words the quantum Zeno effect arises because
$Q^{\Psi}$ is prevented from contributing to the process.

The short time behaviour we have discussed in this paper should not be confused with a discussion of the dominance of classical mechanics in the world we see around us.  Classical phenomena can arise in spite of the presence of a quantum potential energy, provided the kinetic energy is much larger, in which case the effect of this potential can be neglected.   This can be seen by examining equation  (\ref{eq:QHJ}) which clearly reduces to the classical Hamilton-Jacobi equation if the quantum potential energy term is small enough.  

The magnitude of the quantum potential energy also accounts for the fragility of entangled states, where the non-local behaviour of these states is completely accounted for by the quantum potential energy.  The exception to this fragility arises  when an entangled state produces a stable state such as occurs in, for example, the electron pair in the ground state of a helium atom.  In this case, a significant amount of energy is needed to overcome the quantum potential energy.  This is in contrast to an EPR entangled pair of photons where great care must be taken to protect the entangled pairs from environmental effects. (See for example, Xiao-Song {\em et al}.~\cite{xsaz12} )

\section{Acknowledgments.}

Maurice de Gosson has been supported by a research grant from the Austrian
Research Agency FWF (Projektnummer P27773--N13). Basil J. Hiley would like to
thank the Fetzer Franklin Fund of the John E. Fetzer Memorial Trust for their
support. Eliahu Cohen was supported by ERC AdG NLST.

\end{document}